\documentclass[twocolumn,amsmath,amssymb,prb]{revtex4}
\usepackage{graphicx}% Include figure files
\usepackage{dcolumn}% Align table columns on decimal point
\usepackage{bm}% bold math

\begin{document}

\preprint{PRL}
%\preprint{APS/123-QED}

\title{Excitation spectroscopy of single quantum dots at tunable positive, neutral and negative charge states}% Force line breaks with \\

%Optical spectroscopy of single quantum dots at tunable positive, neutral and negative charge states" Phys. Rev B 64, 165301 (2001).

\author{Y. Benny}
\email{byael@tx.technion.ac.il} \affiliation{The Physics Department
and the Solid State Institute, Technion -- Israel Institute of
Technology, Haifa 32000, Israel.}

\author{Y. Kodriano}
\affiliation{The Physics Department and the Solid State Institute,
Technion -- Israel Institute of Technology, Haifa 32000, Israel.}

\author{E. Poem}
\affiliation{The Physics Department and the Solid State Institute, Technion -- Israel Institute of Technology, Haifa 32000, Israel.}

\author{T.~A.~Truong}
\affiliation{Materials Department, University of California, Santa
Barbara, California 93106, USA}

\author{P.~M.~Petroff}
\affiliation{Materials Department, University of California, Santa
Barbara, California 93106, USA}

\author{D. Gershoni}
\affiliation{The Physics Department and the Solid State Institute, Technion -- Israel Institute of Technology, Haifa 32000, Israel.}%Lines break automatically or can be forced with \\

\date{\today}% It is always \today, today,

\begin{abstract}
We present a comprehensive study of the optical transitions and
selection rules of variably charged single self-assembled InAs/GaAs
quantum dots. We apply high resolution polarization sensitive
photoluminescence excitation spectroscopy to the same quantum dot
for three different charge states: neutral and negatively or
positively charged by one additional electron or hole. From the
detailed analysis of the excitation spectra, a full understanding of
the single-carrier energy levels and the interactions between
carriers in these levels is extracted for the first time.
\end{abstract}

\pacs{Valid PACS appear here}% PACS, the Physics and Astronomy
\maketitle

\section{\label{sec:Introduction}Introduction}
The discrete nature of the energy levels in semiconductor
quantum-dots (QDs) has been a subject of many
studies~\cite{Takagahara00,Akimov05,Poem07,Ediger07,Warming09,Kodriano10,Benny11}.
This ``atomic-like" spectral feature of QDs, together with their
compatibility with modern semiconductor-based microelectronics and
optoelectronics, make QD-based devices particularly promising as
building blocks for future technologies involving quantum
information processing (QIP).~\cite{Zanardi98,Imamoglu99,Press10}
Devices that emit single and entangled photons on
demand~\cite{Akopian06,Martini96,Michler00,Benson00,Fattal04} are
typical examples of these potential applications. One important
reason which makes QDs particularly attractive for these
applications is that, unlike atoms, their charge state can be easily
controlled by external fields.~\cite{Couto11,Warburton00,Heiss09}

A detailed understanding of the energy levels of confined carriers in
these QDs, and the interactions between them, is essential for
implementing these potential applications. Therefore, there are many
experimental and theoretical studies aiming at achieving this
goal.~\cite{Beranco95,Dekel0061,Poem07,Ediger07} Various
experimental methods are applied in these studies: polarization
sensitive photoluminescence (PL)~\cite{Poem07,Ediger07,Poem10,Kodriano10}, PL in the presence of electric~\cite{Poem07,Ediger07} and magnetic fields~\cite{Braitbart06}, and PL excitation
(PLE)~\cite{Warming09,Siebert09,Benny11} spectroscopies, second
order intensity correlation measurements~\cite{Kodriano10,Poem10}
and time resolved spectroscopy~\cite{Benny10,Poem10}, just to
mention a few. These various experimental methods, when combined with
many body models, lead to a relatively good understanding of the
emission spectrum of neutral~\cite{Kodriano10,Poem07,Ediger07} and
charged~\cite{Poem10,Poem07,Ediger07} QDs. In a recent study, we
applied one and two photon PLE spectroscopies to a neutral QD to
fully reveal its excitonic and biexcitonic resonance-rich
spectrum.~\cite{Benny11}

In the current work, we use our ability to optically control the QD
charge state in order to apply high spectral resolution polarization
sensitive PLE spectroscopy of the same QD at different charge states. A detailed
understanding of the spectrum of a neutral QD~\cite{Benny11}
leads in turn to an understanding of the spectrum when the QD is charged with
an additional, single electron or heavy-hole. This way, a
comprehensive understanding of single electron and single hole
states and energy levels is achieved, together with
characterizations of the interactions between carriers in these
levels.

\section{\label{sec:Theory}Theory and Experiment}
\subsection{\label{sec:ExpSetup}The experimental setup}
The sample used in this work was grown by molecular-beam epitaxy
(MBE) on a (001) oriented GaAs substrate. One layer of
strain-induced $\rm In_{x}Ga_{1-x}As$ QDs was deposited in the
center of a one wavelength microcavity.  The microcavity was
designed to have a cavity mode which matches the QD emission due to
ground state e-h pair recombinations. Such microcavity significantly
improves photon collection efficiency from emission lines which
resonate with the cavity mode since their light is emitted at normal
incidence. The light from higher energy lines is emitted at energy
dependent angle. Therefore, the spectral window in which light
emitted from within the cavity can be efficiently collected is
defined by the numerical aperture (NA) of the collecting
optics.~\cite{Weisbuch98} Planar microcavity, however, affects only
marginally the radiative rate of emission lines (a few percents
Purcell effect) and therefore the intensity of various emission
lines within the allowed spectral window is not influenced by the
cavity. Absorption resonances measured by PLE spectroscopy are
typically above this window, in a spectral range where the upper DBR
mirror has high reflectivity (about 99 percent). Thus, laser light
coupling is equally inefficient for all the absorption lines within
the stop band of the DBR mirror.

During the growth of the QD layer, the sample was not rotated,
resulting in a gradient in the grown QD density. The estimated QD
density in the sample areas that were measured is $10^8$ $\rm
cm^{-2}$. However, the density of QDs that emit in resonance with
the microcavity mode is more than two orders of magnitude
lower.~\cite{Ramon06} Thus, single QDs separated by a few tens of
micrometers were easily located by scanning the sample surface
during PL measurements. Strong anti-bunching in intensity
auto-correlation measurements was then used to verify that the
isolated QDs are single dots and that they are single photon
sources.

The sample was placed inside a sealed metal tube immersed in liquid
helium, maintaining a temperature of 4.2K. A $\times$60 microscope
objective with a numerical aperture of 0.85 was placed above the
sample and used to focus the excitation laser on the sample surface
and to collect the emitted PL. The relatively high NA of our system
resulted in $\sim$15 meV broad spectral window above the cavity mode
energy in which PL emission from the QDs was efficiently collected.
We used a cw tunable Ti:sapphire laser to scan the energy of the optical excitation. The
laser emission energy could be continuously changed using
coordinated rotations of a three plate birefringent filter and a
thin etalon. The excitation intensity was about $0.5 \mu$Watt,
in which no power broadening of the absorption resonances, was
observed, but the PL emission was reasonably intense, allowing
efficient PLE spectroscopy. The polarization of the light was
adjusted and analyzed using a polarized beam splitter (PBS) and two
pairs of computer-controlled liquid crystal variable retarders
(LCVRs). The PL was spectrally resolved by a 1-meter monochromator
and detected by a cooled CCD camera.~\cite{Benny11}

For the polarization sensitive PLE spectroscopy, we monitored the
polarized emission from an identified PL spectral line while varying
the energy and polarization of the exciting light source. From the
measured variations in the intensity of the emitted PL we construct
the PLE spectrum. Resonances in the PLE spectrum are due to many
carrier states in which a photon is absorbed by
generating an additional electron-hole pair. Increased absorption
results in increased emission intensity from the respective PL lines due to
recombination of an electron-hole pair from a many carrier state to
which the excited carriers are relaxed prior to their recombination.
The polarization sensitivity is due to one to one correspondence between
the many carriers' spin wavefunction and the polarization of the
absorbed or emitted photon.~\cite{Akimov05,Benny11} The energy of a particular
resonance, its relative intensity, the particular lines that it
relaxes to, and its polarization selection rules are then used to
unambiguously identify the many-carrier states which form this
resonance.~\cite{Finley01,Ware05,Benny11}

Variation in the QD charge state was achieved by additional
excitation with very weak intensity light with an energy above the
bandgap. We discovered that the average charge state of the QD
strongly depends on the energy of this minute amount of light. By
changing the light color from red (633 nm) to violet (458 nm) we
succeeded to vary the QD charge state from positive to negative,
respectively. For each color, the average charge was different, as
judged by the intensity of the emission from various charged exciton
states. By mixing two colors, any desired average charge state can
be obtained. While the exact mechanism of this charging control
method is not accurately known, it can be qualitatively explained in
terms of deionization of ionized impurity centers in the vicinity of
the QD.~\cite{Hartmann00,Regelman02} A more quantitative description
of this controlled charging is currently under study, however, it is
beyond the scope of this work.

It is important to note that the charge state of the QD achieved by
this weak high energy illumination is an average state only.
Typically, we observe emission from at least two, more often three
different charge states at a given steady state illumination
condition. Clearly, under these conditions the charge state of the QD
fluctuates in time. The fluctuation times are excitation intensity
dependent and can be straight forwardly measured using second order
intensity correlation measurements. We recently reported on
characteristic optical charging times of few
nanoseconds~\cite{Poem102} and on a mechanism of charge fluctuations
via the dark exciton state. In the current work, the intensity of
the high energy excitation was much lower, resulting in at least an
order of magnitude longer charging times. Since the PL accumulation
times were typically about a second, charge fluctuations times are
still too short to cause observable ``telegraphic" noise in the
measurements.

\subsection{\label{sec:PLTrion}Photoluminescence spectra}
\begin{figure*}
\includegraphics[width=0.98\textwidth]{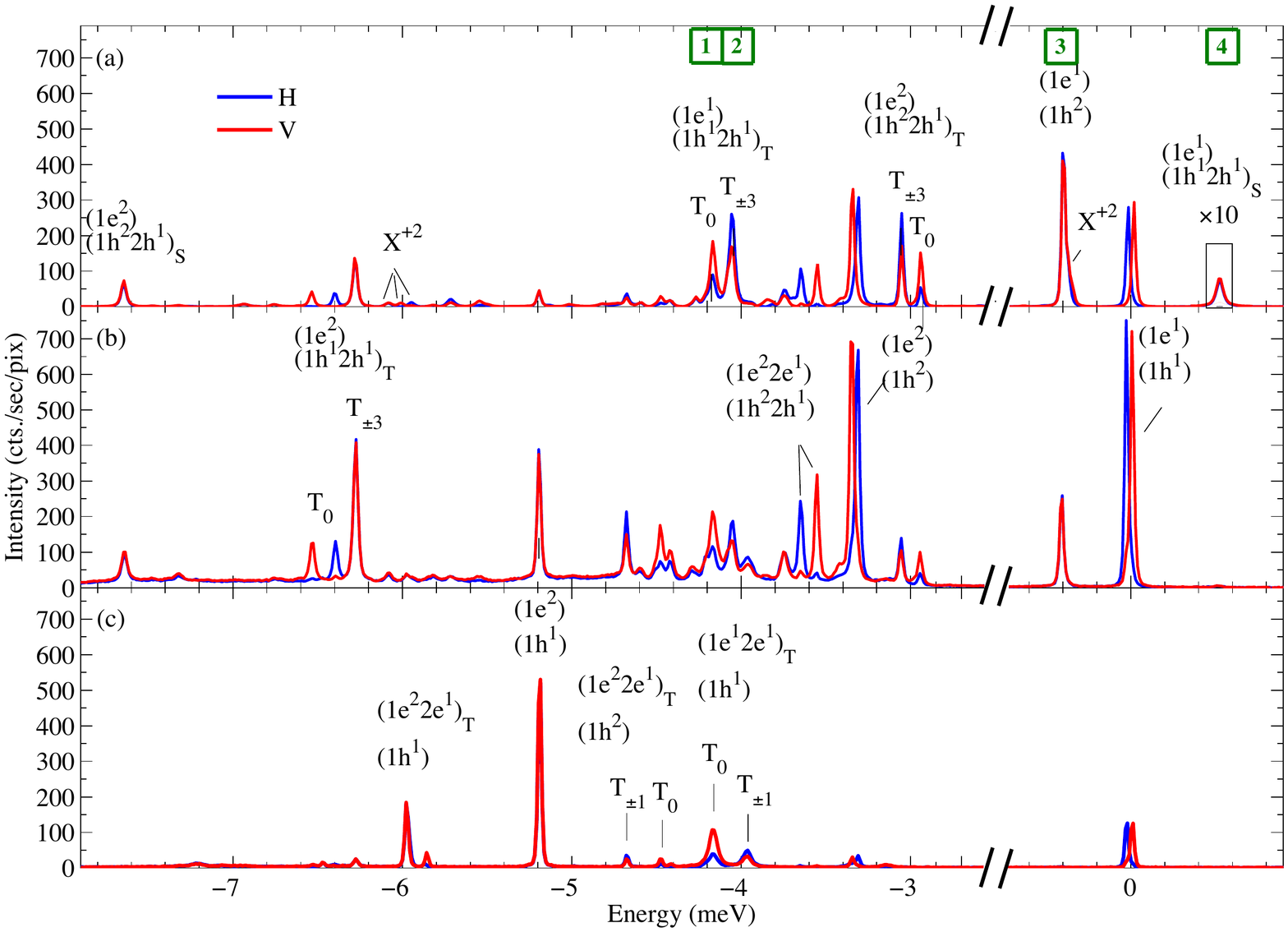}
\caption{\label{fig:PL} Rectilinearly polarized horizontal (blue
line) and vertical (red line) PL spectra of the single QD, at
various charge states. In (a) the  QD is on average positively
charged by an additional heavy hole, in (b) it is neutral, and in
(c) it is charged negatively, having on average one additional
electron. The spectral lines due to excitonic and biexcitonic
recombinations are identified in the Figure, denoted by the initial
state of the optical transition. The numbers in bold rectangles
in panel (a) denote optical transitions with the same numbers as in
Fig.\ \ref{fig:EnergyDiagram}.}
\end{figure*}

Fig.\ \ref{fig:PL} presents rectilinear horizontal (H) and vertical
(V) polarized PL spectra of a single QD in resonance with the
microcavity mode for three different average charge states. The
observed  spectral lines are identified in the Figure.

The notation used by Benny \emph{et al.}~\cite{Benny11} is used in this work as well: A single
carrier state is described by its envelope wavefunction or orbital
mode (O=1,2,...,6), where the number represents the energy order of
the level, so that O=1 represents the ground state. O is followed by
the type of carrier, electron (e) or heavy-hole (h) and a
superscript which describes the occupation of the single carrier
state. The superscript can be either 1 (open shell) or 2 (closed
shell), subject to the Pauli exclusion principle (unoccupied
states are not included in the description). All the occupied states
of carriers of same type are then marked by subscripts which
describe the mutual spin configuration ($\sigma$) of these states. A
full description of a positive trion with two unpaired holes has
therefore the form
$\rm(O_{e_1}e^{1})_{\sigma_e}(O_{h_1}h^{1}O_{h_2}h^{1})_{\sigma_h}$
and in the same way a negative trion with two unpaired electrons has
the form
$\rm(O_{e_1}e^{1}O_{e_2}e^{1})_{\sigma_e}(O_{h_1}h^{1})_{\sigma_h}$.

Fig.\ \ref{fig:PL}(a) presents the PL emission spectrum of the QD populated on average with one positive charge. This PL was obtained while the
QD was excited by a HeNe laser light. The spectrum in this case is
dominated by an unpolarized spectral line which appears 0.4 meV
below the neutral exciton line. Based on previous studies, we
unambiguously identify this spectral line as the optical transition from
the ground state positive trion ($X^{+1}$),
$(1e^1)(1h^2)\rightarrow(1h^1)$ to a single ground level heavy hole
state.~\cite{Poem10} In the initial state, the two holes form a singlet of
spin 0 in their ground level. The total spin of this state is
therefore given by the  spin-half state of the single-electron. As
expected, this total half-integer spin state is doubly (Kramers)
degenerate in the absence of external time reversal breaking
perturbation, such as a magnetic field.

The selection rules for optical transitions from this state are
determined by the spin of the recombining
electron.~\cite{Kalevich05} A spin-up, $\uparrow$(-down,
$\downarrow$) electron recombines with a spin-down, $\Downarrow$
(-up, $\Uparrow$) hole
 to give a left (right) hand circularly polarized
photon with angular momentum of -1 (+1).
Since excitation at this high, above bandgap energy,
photogenerates an equal mixture of the two electron spin states, the
PL that results from the recombination is unpolarized.

The energy separation between the first and second hole levels is
smaller than that between the corresponding electron levels, mainly
due to the hole's larger effective mass. The lowest energy excited
level of the positive trion is therefore one in which one of the
holes is in its first excited level.~\cite{Poem10} The total spin of
the two holes that occupy the QD is $3$, and thus there are four possible
states with three possible spin projections for the two hole total
spin; $+3$, ($-3$) when both holes are aligned with spin up
(down), and two states with total spin projection $0$, when the holes
spins are anti-aligned. The hole-hole exchange interaction removes
the degeneracy between the singlet state, $1/\sqrt{2}(\Downarrow\Uparrow-\Uparrow\Downarrow)$
($S^h$), which is antisymmetric under carrier exchange, and the three, lower energy, symmetrical states,
$\Downarrow\Downarrow$ ($T_{-3}^h$), $\Uparrow\Uparrow$ ($T_{+3}^h$) and
$1/\sqrt{2}(\Downarrow\Uparrow+\Uparrow\Downarrow)$ ($T_{0}^h$).
There is experimental evidence, based on recombination of the
doubly positively charged exciton, that the degeneracy of the triplet states is further removed to
a higher energy singlet with spin projection $0$ and a lower energy
doublet with spin projection $\pm3$.~\cite{Ediger07}
Here, in the case of the excited positive trion, this degeneracy is,
in any case, removed by the electron-hole exchange interaction with
the single electron.

The effect of the electron-hole exchange
interactions on the triplet states of a trion are well described
elsewhere in terms of the pseudospin
Hamiltonian.~\cite{Akimov05,Kavokin03}
Fig.\ \ref{fig:EnergyDiagram} describes the energies of these eigenstates.
The emission and absorption transitions of the positively charged
trion are described in the Figure as well. For
simplicity, these transitions are denoted by numbered arrows where
the numbers reflect their energy order and up (down) arrow
symbolizes photon absorption (emission). In a simple model, optical transitions between electron and hole
levels of different orbital symmetries are forbidden, due to zero
overlap between the electron and hole envelope
wavefunctions.~\cite{Poem07} However, our experimental results
clearly indicate that these transitions are allowed, probably due to
symmetry breaking between the electron and hole
potentials.~\cite{Warming09,Siebert09}

We note in Fig.\ \ref{fig:EnergyDiagram} that the optically allowed transitions from the triplet
states are partially rectilinearly polarized, due to the mixing between the $T_{\pm3}^h$ and $T_{0}^h$ states, described by the coefficients $\alpha$ and $\beta$.~\cite{Akimov05,Kavokin03} These two transitions
are identified in the PL spectrum, where the transition from the
$T_{\pm3}^h$ state appears at -4.06 meV  and the transition from the
$T_{0}^h$ state appears at -4.17 meV.~\cite{Poem10}
 At an energy of 0.43 meV, about 4.6 meV higher, the transition from the singlet $(1e^1)(1h^12h^1)_S\rightarrow(2h^1)$ is identified.
The transitions from the positive ground state biexciton,
$(1e^2)(1h^22h^1)$, to these triplet states are identified in the
spectrum as well. These biexcitonic transitions were identified
previously by time-resolved intensity correlation
measurements.~\cite{Poem10} Additional lines are identified as transitions related to a two-hole charged QD ($X^{+2}$). These transitions present exactly the same resonances in their PLE spectra (not shown here), and thus we attribute them to emission lines that occur from the same initial state.

Fig.\ \ref{fig:PL}(b) presents the PL spectrum excited with 488 nm
Ar+ laser light. Under these conditions the QD is on average
neutral. One clearly notes that neutral optical transitions in this
spectrum are stronger than transitions in the presence of additional
charge. The ground state neutral exciton and biexciton are readily
identified by their well studied fine-structure cross-rectilinearly
polarized doublets.~\cite{Ivchenko, Gammon96} The biexciton doublet,
about 3.3 meV lower in energy than the exciton doublet, has the same
energy separation between its two cross-rectilinearly polarized
components as the exciton albeit with reversed order, as
expected.~\cite{Poem07,Benny11} Additional neutral transitions are
observed, resulting from the spin-blockaded hole triplet states of
the first excited biexciton, where one hole is in the second orbital
energy mode. These lines are discussed in detail
elsewhere.~\cite{Kodriano10,Benny11} The transitions from the neutral triexciton, where the QD is populated by three electrons and three holes, $(1e^22e^1)(1h^22h^1)$, to the neutral electron-triplet-hole-triplet biexciton states~\cite{Benny11} are identified in this spectrum as well.

Fig.\ \ref{fig:PL}(c) presents the PL spectrum excited by 458 nm Ar+
laser light. Under these excitation conditions, the QD is on average
negatively charged with one electron. The optical transitions in
which the QD is negatively charged become much stronger than the
neutral transitions, while positive transitions nearly vanish. The
eigenstates of the negatively charged QD can be described in a
similar way to those of the positively charged one. Here, the PL
spectrum is dominated by an unpolarized spectral line due to
recombination from the ground state of the negative trion,
$(1e^2)(1h^1)\rightarrow(1e^1)$.  In addition, the transitions from
the metastable, spin-blockaded excited trion levels, where one
electron occupies the first excited electron orbital level, are
observed. Similarly to the positive trion, four different spin
configurations are expected; one anti-symmetric with respect to
electron exchange with total spin projection $0$,
$1/\sqrt{2}(\uparrow\downarrow-\downarrow\uparrow)$ ($\rm S^e$), and
three symmetric triplet states with total spin projection $0$,
$1/\sqrt{2}(\uparrow\downarrow+\downarrow\uparrow)$ ($\rm T_{0}^e$),
and  1 (-1) : $\uparrow\uparrow$
($T_{+1}^e$) [$\downarrow\downarrow$ ($T_{-1}^e$)]. Due to the
electron-electron exchange interaction the singlet state is higher
in energy by a few meV, and the electron-hole exchange interaction
further removes the degeneracy between the triplet states.

The transitions from the ground state of the negatively charged
biexciton to the metastable spin-blockaded triplet states of the
exciton are also observed in the PL spectrum of Fig.\
\ref{fig:PL}(c). The identification of these biexcitonic transitions
was confirmed by time-resolved intensity correlation measurements
between the cascading transitions in a similar manner to the
positive transitions~\cite{Poem10} (not shown here). The intense
line at about -6 meV is observed only in the PL spectrum of the
negatively charged QD state. We attribute this line to a transition
from a doubly negatively charged QD
$(1e^22e^1)(1h^1)\rightarrow(1e^12e^1)_T$ ($X^{-2}$).

\subsection{\label{sec:PLETrion}Photoluminescence excitation spectra}
\begin{figure}
\includegraphics[width=0.5\textwidth]{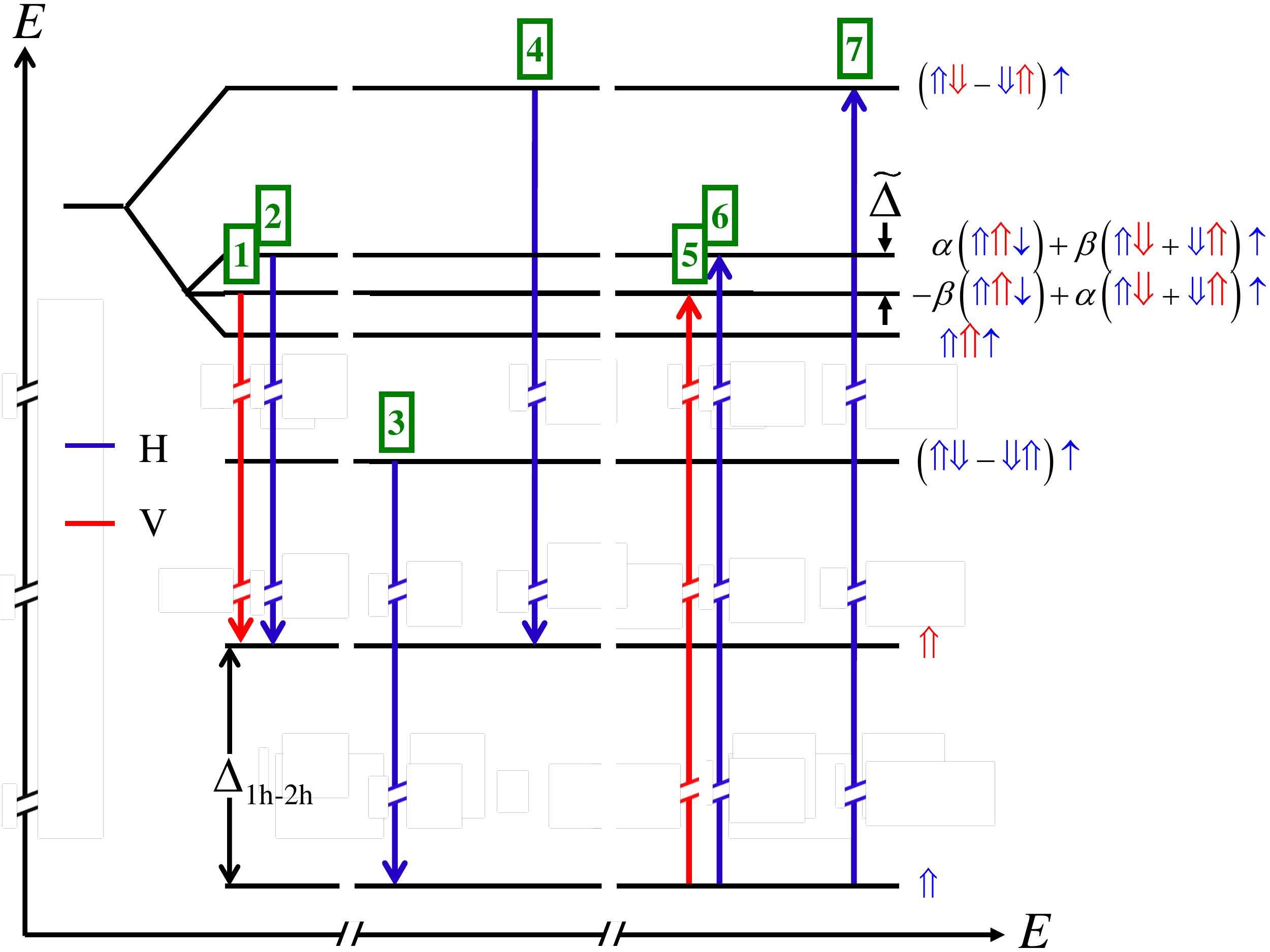}
\caption{\label{fig:EnergyDiagram} Energy diagram of the positively
charged trion states, $(1e^1)(1h^2)$ and $(1e^1)(1h^12h^1)$ and the
optical transitions to and from these states.  Blue (red) arrows
represent partial H (V) linear polarization.~\cite{Akimov05} The
optical transitions are numbered in increasing energy order, and the
numbers corresponds to transitions observed in PL (Fig.1) and PLE
(Fig.4) spectra. The spin configurations are presented to the right.
Blue single (double) arrows represent a ground state electron (hole)
and red single (double) arrow is for an excited electron (hole).
$\alpha$ and $\beta$ are the coefficients of mixing between the
$T_{\pm3}$ and $T_{0}$ states.~\cite{Akimov05}}
\end{figure}

\begin{figure}
\includegraphics[width=0.48\textwidth]{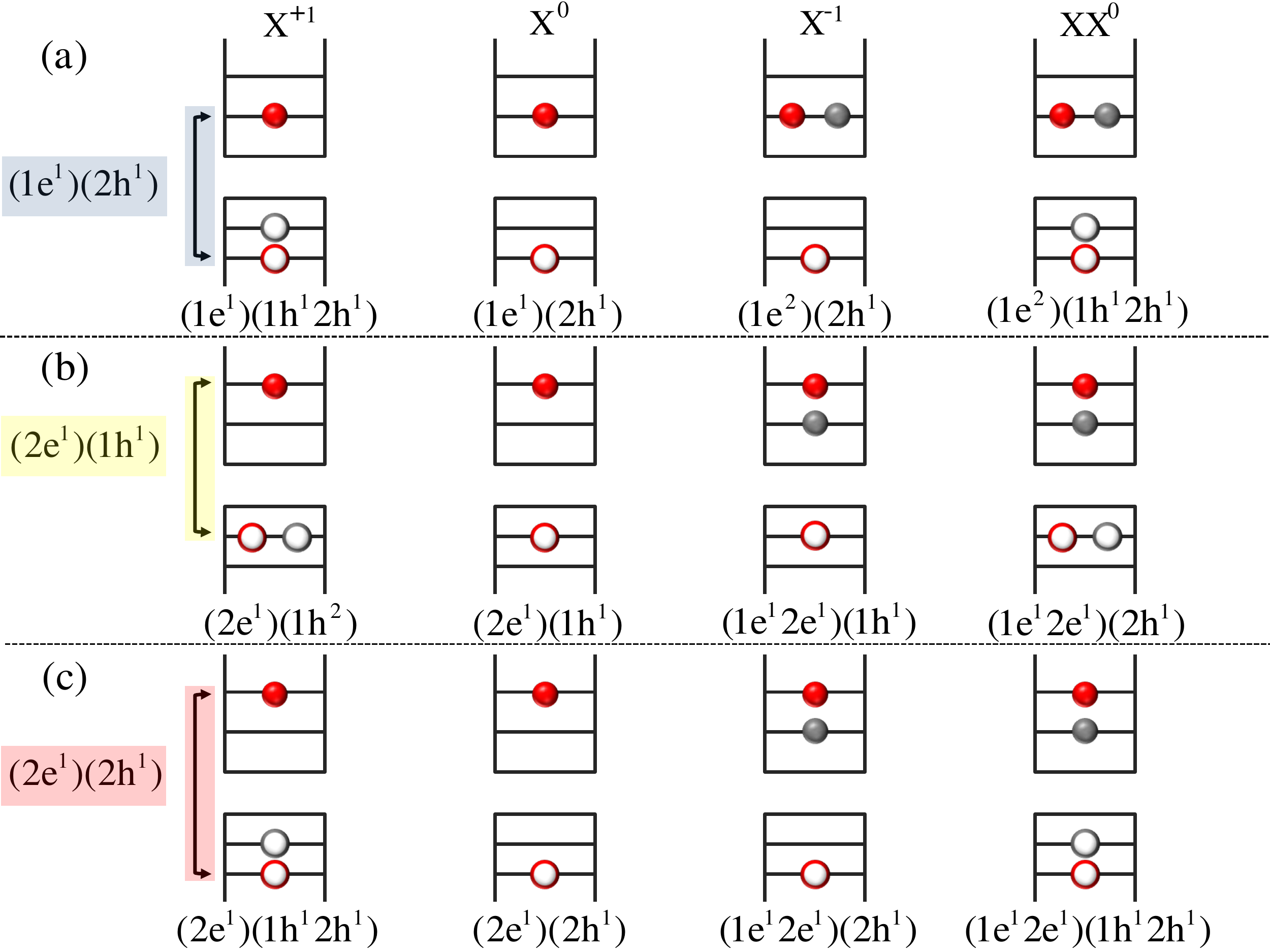}
\caption{\label{fig:StatesPLE} Schematic descriptions of the
multi-carrier resonances observed in the PLE spectra (Fig.\
\ref{fig:PLE}). Optical transitions in which $1e^1-2h^1$,
$2e^1-1h^1$ and $2e^1-2h^1$ electron-hole pair are photogenerated
are described in (a), (b) and (c), respectively. Transitions
involving the positive trion, the neutral exciton, the negative
trion and the neutral biexciton are presented, in the first, second
, third and fourth column, respectively. The background colors are
used for identifying these transitions in Fig.\ \ref{fig:PLE}.
Filled (empty) red circle represents the photogenerated electron
(hole) and a filled (empty) gray circle represents a resident
electron (hole).}
\end{figure}

\begin{figure*}
\includegraphics[width=0.97\textwidth]{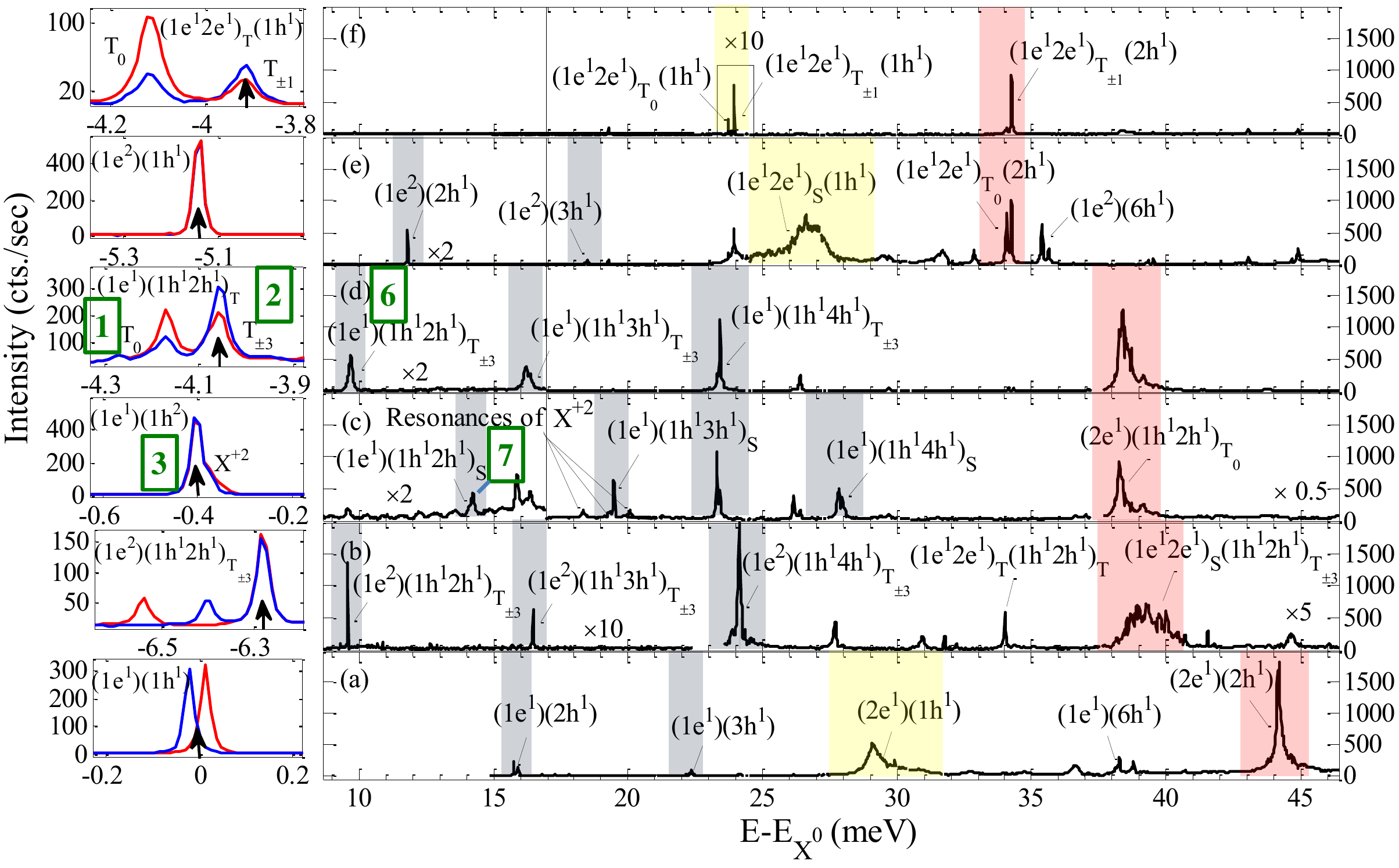}
\caption{\label{fig:PLE} Measured linear horizontal (blue) and
vertical (red) polarized PL (left panels) and total linear polarized
PLE (right panels) spectra of the transitions
$(1e^1)(1h^1)\rightarrow0$ (a),
$(1e^1)(1h^12h^1)_{T_{\pm3}}\rightarrow(1e^1)(2h^1)$ (b),
$(1e^1)(1h^2)\rightarrow(1h^1)$ (c)
$(1e^1)(1h^12h^1)_{T_{\pm3}}\rightarrow(2h^1)$ (d),
$(1e^2)(1h^1)\rightarrow(1e^1)$ (e) and
$(1e^12e^1)_{T_{\pm1}}(1h^1)\rightarrow(2e^1)$ (f). The assignment
of the resonances are marked on the figures, above the observed
resonances. The numbers in bold rectangles in panels (c) and (d)
denote optical transitions with the same numbers in Fig.\
\ref{fig:EnergyDiagram}.}
\end{figure*}

\begin{figure}
\includegraphics[width=0.48\textwidth]{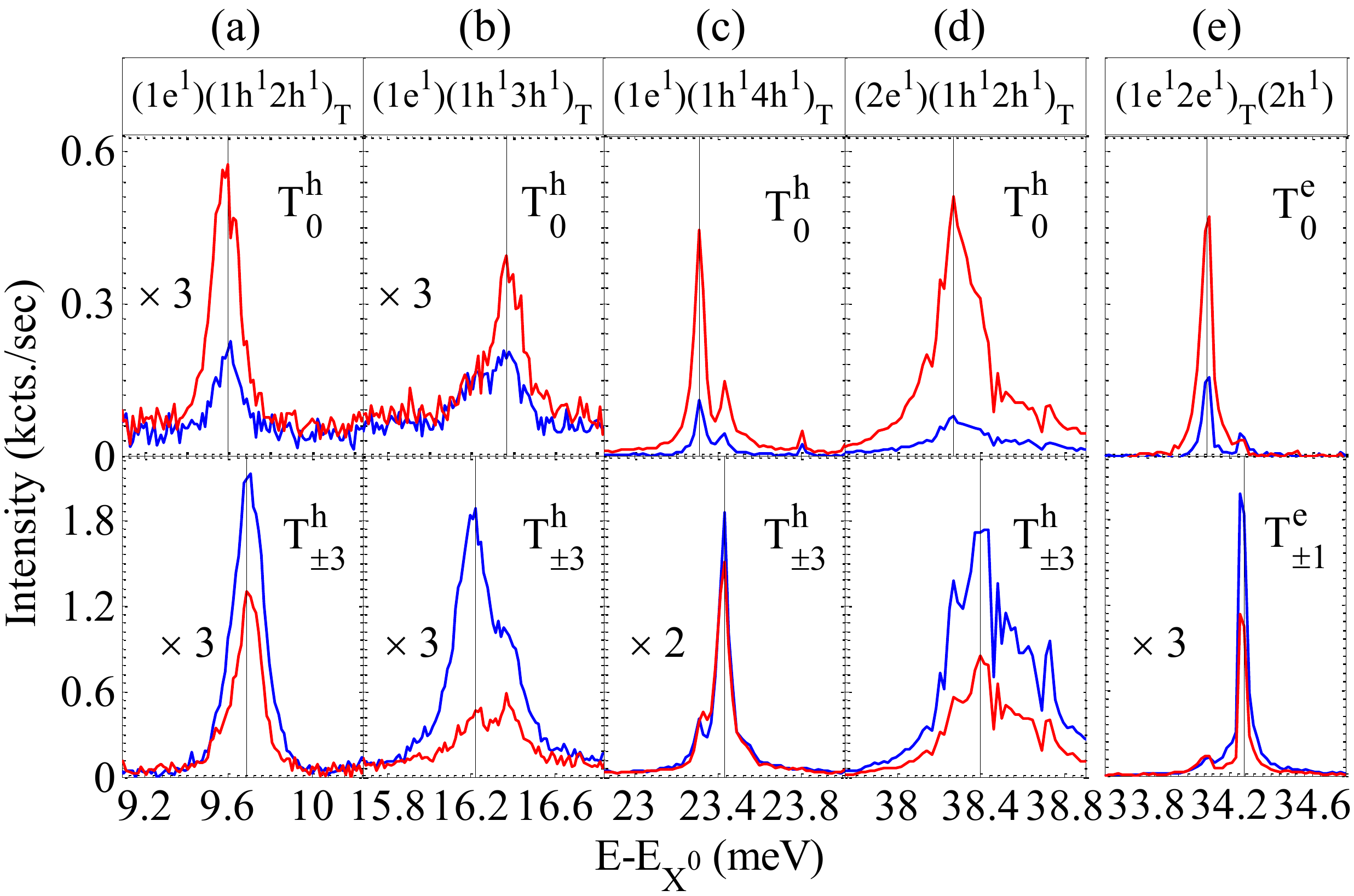}
\caption{\label{fig:PLE_Polarized} HH (blue line) and VV (red line)
rectilinearly polarized PLE spectra, where the first (second) letter
indicates the polarization of the exciting (emitted) light. The
following resonances are presented: (a)
$(1e^1)(1h^12h^1)_{T}$, (b) $(1e^1)(1h^13h^1)_{T}$,
(c) $(1e^1)(1h^14h^1)_{T}$, (d)
$(2e^1)(1h^12h^1)_{T}$, (e) $(1e^12e^1)_{T}(2h^1)$.}
\end{figure}

Higher energy transitions were studied by PLE measurements.
The main resonances in the PLE can be easily
understood by inspecting the simple state diagrams presented in
Fig.\ \ref{fig:StatesPLE}. In these diagrams the initial carriers
are presented in grey and the resonantly photogenerated
electron-hole pair is presented in red. Each row in Fig.\ \ref{fig:StatesPLE}
corresponds to a certain photoexcited pair, and each column corresponds to
a particular QD charge occupation.

The first row of Fig.\ \ref{fig:StatesPLE}(a) describes various transitions
in which a ground-state electron and a first-excited-state hole,
$(1e^1)(2h^1)$, are photogenerated. If the pair is
added to an empty QD, an exciton will be formed ($X^{0}$). If the QD
is populated by a single electron, the added pair will result in a
negative trion ($X^{-1}$) in which the two electrons are paired in
the ground state (singlet). If the QD is populated by a hole, four
different Kramers degenerate spin-configurations of a positively
charged trion ($X^{+1}$) may be formed: three metastable
spin-blockaded states in which the holes form triplets, and one state in
which the holes form a singlet state. If the QD is occupied by an
exciton, then the added pair will result in one of the neutral
biexcitonic ($XX^{0}$) e-singlet-h-triplet states. The transitions
described in this row are marked blue.

In the second row, Fig.\ \ref{fig:StatesPLE}(b), the photogenrated
electron is in its second energy level and the photogenerated hole
is in its ground state $(2e^1)(1h^1)$. In general,
these resonances are expected at higher energies than the
corresponding resonances (same column) in Fig.
\ref{fig:StatesPLE}(a), because the separation between the energy
levels of the electrons is larger than that of the heavy holes.
There is, however, one important difference between the two cases.
The energy difference between the first two electron levels closely
resonates with the energy of one longitudinal optical (LO) phonon in
the materials composing the QD and the wetting layer. This results
in efficient LO phonon mediated coupling between these electronic
levels. This coupling has important consequences on the observed
spectra, as previously noted~\cite{Hameau99,Benny11} and further
discussed below. The transitions described in this row are marked
yellow.

In the third row, Fig.\ \ref{fig:StatesPLE}(c), both photogenerated
carriers, the electron and hole, are in their first excited state,
$(2e^1)(2h^1)$. Thus, these transitions are at higher energies than
the transitions in rows (a) and (b). The trions in this case have
four Kramers' degenerate spin configurations of the majority
carriers, as in the case of rows (a)-(b); three triplet states and
one singlet state. One of the four levels, that in which all
the carriers spins are aligned, is optically inaccessible. The
neutral biexciton, however, has sixteen spin configurations ($2^4$):
e-triplet-h-triplet (nine states), e-triplet-h-singlet (3 states),
e-singlet-h-triplet (3 states) and e-singlet-h-singlet (1 state).
Here as well, two states in which all the carriers' spins are
aligned are optically inaccessible.~\cite{Benny11} The LO phonon
mediated coupling between the first two electronic levels, has
similar signature in the PLE spectra from Fig.\ \ref{fig:StatesPLE}(c) as it has in the
PLE spectra of Fig.\ \ref{fig:StatesPLE}(b). The transitions described in this row are
red.

In Fig.\ \ref{fig:PLE}, we present the PLE spectra of the main optical
transitions presented in Fig.\ \ref{fig:PL}, together with a high
resolution polarization sensitive PL spectra of the spectral lines
used for monitoring the PLE spectra.

Fig.\ \ref{fig:PLE}(a) and Fig.\ \ref{fig:PLE}(b) present the PLE
spectra of the neutral exciton and biexciton, respectively. These
spectra were presented and discussed previously.~\cite{Benny11} Here
they are displayed for comparison with the PLE spectra of the
charged trions. The PLE spectra of the ground and first excited positive trion are presented in Fig.\
\ref{fig:PLE}(c) and Fig.\ \ref{fig:PLE}(d), respectively. Similarly, the
PLE from the ground and first excited negative trion
are presented in Fig.\ \ref{fig:PLE}(e) and Fig.\ \ref{fig:PLE}(f),
respectively. The identified resonances are denoted in each
spectrum.

The transitions that we describe in Fig.\ \ref{fig:StatesPLE} are
clearly identified in the  PLE spectra in Fig.\ \ref{fig:PLE}(a-f),
where they are colored according to their classification. We note
that transitions in which the hole is excited to its 2nd, 3rd and
4th energy levels are similar in nature. Therefore they are all
marked blue in Fig.\ \ref{fig:PLE}. The identifications of the
various lines are based here on their energetic order, their
oscillator strength, and the emission line which monitored their
absorption. In addition we utilized  the similarities between the
spectra from charged states to that from the neutral
state~\cite{Benny11} of the same QD. We show below that polarization
sensitive PLE measurements further support our line identifications.

By inspecting Fig.\ \ref{fig:EnergyDiagram} one immediately sees that
the energy difference between the first and second hole orbitals,
$\Delta_{1h-2h}$, can be directly extracted from the energy
difference between the absorption transitions 5,6 and 7
[$(1h^1)\rightarrow(1e^1)(1h^12h^1)_{\sigma_h}$] and the emission
transitions 1,2 and 4,
[$(1e^1)(1h^12h^1)_{\sigma_h}\rightarrow(2h^1)$], respectively. Thus, using Figs.\ \ref{fig:PLE}(c,d),
[transition 5 is seen in the PLE spectrum of the PL from the line
$(1e^1)(1h^12h^1)_{T_0}\rightarrow(2h^1)$ which is not shown here]  and Fig.\
\ref{fig:PL} we obtain $\Delta_{1h-2h}=13.70\pm0.02$ meV.

Similarly, the difference between the first and second electron
orbitals can be determined by inspecting Fig.\ \ref{fig:PLE}(f). One
finds that the energy difference between the doublet in the PLE
spectrum due to the optical transitions
$(1e^1)\rightarrow(1e^12e^1)_{T_0}(1h^1)$ and
$(1e^1)\rightarrow(1e^12e^1)_{T_{\pm1}}(1h^1)$ and the doublet in PL due to the
optical transitions $(1e^12e^1)_{T_0}(1h^1)\rightarrow(2e^1)$ and
$(1e^12e^1)_{T_{\pm1}}(1h^1)\rightarrow(2e^1)$ is exactly
$\Delta_{1e-2e}=27.85\pm0.02$ meV.

We note that the first excited state of the negative trion is a
singlet [see Fig.\ \ref{fig:PLE}(e)] while the first one of the
positive trion is a triplet [see Fig.\ \ref{fig:PLE}(d)]. This
difference is expected since the lowest energy excited resonance is
obtained by promoting a heavy hole, rather than an electron (Fig.\
\ref{fig:StatesPLE}). This further supports our spectroscopic
identification.

The symmetry-allowed optical transitions between the second
electron and hole states described in Fig.\
\ref{fig:StatesPLE}(c) give rise to very strong resonances in the
various PLE spectra (colored red). We note
however that the linewidth of these symmetry-allowed transitions
differ significantly. While the line-width of resonances to states
in which the two electrons form triplets is very narrow, transitions
to states in which there is only single electron in the second
electron state, or to states in which the two electrons form singlets,
are quite broad. We attribute this broadening to optical
phonon-induced strong coupling between the first and second
electronic orbitals. The LO phonon dispersion broadens these
transitions and enhances their optical strength. Since the phonon
does not mix states of different electronic spin, triplet electronic
states are not coupled to the ground singlet state.
We note that these resonances are about 29 meV higher than the
corresponding resonances in which the electron is in the first
level. This energy separation characterizes the energy of LO phonons in compounds of GaAs and InAs.~\cite{Sarkar05,Sarkar08,Lemaitre01,Findeis00}

The phonon induced strong coupling also enables the
symmetry  forbidden optical transition that results in
photogeneration of a $(2e^1)(1h^1)$ electron-hole pair.~\cite{Hameau99}
These transitions become strong and broad as marked yellow in Fig.\ \ref{fig:PLE}(a,e). Their energy
difference is about 29 meV higher than the corresponding transition
into the first electronic orbital.

In Fig.\ \ref{fig:PLE_Polarized}, the rectilinearly polarized PLE
spectra of a few of the resonances observed in Fig.\ \ref{fig:PLE}(d)
and (f) are displayed. The resonances appear in pairs since they
result from the triplet hole states $T_{0}^h$ and $T_{\pm3}^h$ [Fig.\ \ref{fig:PLE_Polarized}(a-d)] and from the triplet electron
states $T_{0}^e$ and $T_{\pm1}^e$ [Fig.\ \ref{fig:PLE_Polarized}(e)].
Spectra in which the exciting light is polarized H (V) and the
detection is polarized H(V) are displayed in blue (red).

We note that, due to the anisotropic electron-hole exchange, the two
triplet states are mixed and therefore partially linearly polarized
as schematically described in Fig.\ \ref{fig:EnergyDiagram}. This
partial linear polarization holds also in Fig.\
\ref{fig:PLE_Polarized}(d) despite the LO-phonon induced
spectral broadening. As mentioned above, phonon mediated transitions
preserve the electronic spin and therefore also preserve the optical
polarization. Additionally, in all the spectra except the one
presented in Fig.\ \ref{fig:PLE_Polarized}(b) the higher energy
transition among the doublet, is polarized H and leads into the
triplet states in which the majority carriers' spins are aligned
($T_{\pm3}^h$ and $T_{\pm1}^e$ for holes and electrons,
respectively). In Fig.\ \ref{fig:PLE_Polarized}(b), however, this
energy order is reversed. We attribute this reversal to the
difference in the symmetry of the excited hole state, which is
$p_V$-like, rather than $p_H$-like or $d_{HH}$-like  in all the other
cases. This symmetry difference results in the sign reversal of
$\widetilde{\Delta}$ in Fig.2.~\cite{Kavokin03,Poem07}

\section{\label{sec:Summary}Summary}
In summary, we present a comprehensive study of the optical
transitions of the same single self assembled quantum dot in various
charge states. Our study provides a systematic way of understanding
the rich photoluminescence and photoluminescence excitation spectra
that such quantum dots reveal and in particular, a direct
measurement of the confined single carriers' energy level
separations. The experimental and theoretical tools that we
developed for the spectroscopic characterization of these variably
charged quantum dots are essential for achieving coherent control of
carrier spins in semiconductor quantum dots and for possible
implementation of quantum logic.

%\newpage
\begin{acknowledgments}
The support of the US-Israel Binational Science Foundation (BSF),
the Israeli Science Foundation (ISF), the Israeli Ministry of
Science and Technology (MOST), Eranet Nano Science Consortium and
that of the Technion's RBNI are gratefully acknowledged.
\end{acknowledgments}


\begin{thebibliography}{40}
\expandafter\ifx\csname natexlab\endcsname\relax\def\natexlab#1{#1}\fi
\expandafter\ifx\csname bibnamefont\endcsname\relax
  \def\bibnamefont#1{#1}\fi
\expandafter\ifx\csname bibfnamefont\endcsname\relax
  \def\bibfnamefont#1{#1}\fi
\expandafter\ifx\csname citenamefont\endcsname\relax
  \def\citenamefont#1{#1}\fi
\expandafter\ifx\csname url\endcsname\relax
  \def\url#1{\texttt{#1}}\fi
\expandafter\ifx\csname urlprefix\endcsname\relax\def\urlprefix{URL }\fi
\providecommand{\bibinfo}[2]{#2}
\providecommand{\eprint}[2][]{\url{#2}}

\bibitem[{\citenamefont{\rm{T.\ Takagahara}}(2000)}]{Takagahara00}
\bibinfo{author}{\bibnamefont{\rm{T.\ Takagahara}}}, \bibinfo{journal}{Phys.\
  Rev.\ B} \textbf{\bibinfo{volume}{62}}, \bibinfo{pages}{16840}
  (\bibinfo{year}{2000}).

\bibitem[{\citenamefont{\rm{I.\ A.\ Akimov, K.\ V.\ Kavokin, A.\ Hundt, F.\
  Henneberger}}(2005)}]{Akimov05}
\bibinfo{author}{\bibnamefont{\rm{I.\ A.\ Akimov, K.\ V.\ Kavokin, A.\ Hundt,
  F.\ Henneberger}}}, \bibinfo{journal}{Phys. Rev. B.}
  \textbf{\bibinfo{volume}{71}}, \bibinfo{pages}{075326}
  (\bibinfo{year}{2005}).

\bibitem[{\citenamefont{{E.\ Poem, J.\ Shemesh, I.\ Marderfeld, D.\ Galushko,
  N.\ Akopian, D.\ Gershoni, B.\ D.\ Gerardot, A.\ Badolato, P.\ M.\
  Petroff}}(2007)}]{Poem07}
\bibinfo{author}{\bibnamefont{{E.\ Poem, J.\ Shemesh, I.\ Marderfeld, D.\
  Galushko, N.\ Akopian, D.\ Gershoni, B.\ D.\ Gerardot, A.\ Badolato, P.\ M.\
  Petroff}}}, \bibinfo{journal}{Phys.\ Rev.\ B} \textbf{\bibinfo{volume}{76}},
  \bibinfo{pages}{235304} (\bibinfo{year}{2007}).

\bibitem[{\citenamefont{{M.\ Ediger, G.\ Bester, A.\ Badolato, P.\ M.\ Petroff,
  K.\ Karrai, A.\ Zunger, R.\ J.\ Warburton}}(2007)}]{Ediger07}
\bibinfo{author}{\bibnamefont{{M.\ Ediger, G.\ Bester, A.\ Badolato, P.\ M.\
  Petroff, K.\ Karrai, A.\ Zunger, R.\ J.\ Warburton}}},
  \bibinfo{journal}{Nature Physics} \textbf{\bibinfo{volume}{3}},
  \bibinfo{pages}{774} (\bibinfo{year}{2007}).

\bibitem[{\citenamefont{{T.\ Warming, E.\ Siebert, A.\ Schliwa, E.\ Stock, R.\
  Zimmermann, D.\ Bimberg}}(2009)}]{Warming09}
\bibinfo{author}{\bibnamefont{{T.\ Warming, E.\ Siebert, A.\ Schliwa, E.\
  Stock, R.\ Zimmermann, D.\ Bimberg}}}, \bibinfo{journal}{Phys.\ Rev.\ B}
  \textbf{\bibinfo{volume}{79}}, \bibinfo{pages}{125316}
  (\bibinfo{year}{2009}).

\bibitem[{\citenamefont{{Y.\ Kodriano, E.\ Poem, N.\ H.\ Lindner, C.\
  Tradonsky, B.\ D.\ Gerardot, P.\ M.\ Petroff, J.\ E.\ Avron, D.\
  Gershoni}}(2010)}]{Kodriano10}
\bibinfo{author}{\bibnamefont{{Y.\ Kodriano, E.\ Poem, N.\ H.\ Lindner, C.\
  Tradonsky, B.\ D.\ Gerardot, P.\ M.\ Petroff, J.\ E.\ Avron, D.\ Gershoni}}},
  \bibinfo{journal}{Phys. Rev. B} \textbf{\bibinfo{volume}{82}},
  \bibinfo{pages}{155329} (\bibinfo{year}{2010}).

\bibitem[{\citenamefont{{Y.\ Benny, Y.\ Kodriano, E.\ Poem, S.\ Khatsevitch,
  D.\ Gershoni, P.\ M.\ Petroff}}(2011)}]{Benny11}
\bibinfo{author}{\bibnamefont{{Y.\ Benny, Y.\ Kodriano, E.\ Poem, S.\
  Khatsevitch, D.\ Gershoni, P.\ M.\ Petroff}}}, \bibinfo{journal}{Phys. Rev.
  B} \textbf{\bibinfo{volume}{84}}, \bibinfo{pages}{075473}
  (\bibinfo{year}{2011}).

\bibitem[{\citenamefont{{P.\ Zanardi, F.\ Rossi}}(1998)}]{Zanardi98}
\bibinfo{author}{\bibnamefont{{P.\ Zanardi, F.\ Rossi}}},
  \bibinfo{journal}{Phys.\ Rev.\ Lett.} \textbf{\bibinfo{volume}{81}},
  \bibinfo{pages}{4752} (\bibinfo{year}{1998}).

\bibitem[{\citenamefont{{A.\ Imamo\={g}lu, D.\ D.\ Awschalom, G.\ Burkard, D.\
  P.\ DiVincenzo, D.\ Loss, M.\ Sherwin, A.\ Small}}(1999)}]{Imamoglu99}
\bibinfo{author}{\bibnamefont{{A.\ Imamo\={g}lu, D.\ D.\ Awschalom, G.\
  Burkard, D.\ P.\ DiVincenzo, D.\ Loss, M.\ Sherwin, A.\ Small}}},
  \bibinfo{journal}{Phys.\ Rev.\ Lett.} \textbf{\bibinfo{volume}{83}},
  \bibinfo{pages}{4204} (\bibinfo{year}{1999}).

\bibitem[{\citenamefont{{D.\ Press, K.\ De Greve, P.\ L.\ McMahon, T.\ D.\
  Ladd, B.\ Friess, C.\ Schneider, M.\ Kamp, S.\ H\"{ö}fling, A.\ Forchel, Y.\
  Yamamoto}}(2010)}]{Press10}
\bibinfo{author}{\bibnamefont{{D.\ Press, K.\ De Greve, P.\ L.\ McMahon, T.\
  D.\ Ladd, B.\ Friess, C.\ Schneider, M.\ Kamp, S.\ H\"{ö}fling, A.\ Forchel,
  Y.\ Yamamoto}}}, \bibinfo{journal}{Nature Photonics}
  \textbf{\bibinfo{volume}{4}}, \bibinfo{pages}{367} (\bibinfo{year}{2010}).

\bibitem[{\citenamefont{\rm{N.\ Akopian, N.\ H.\ Lindner, E.\ Poem, Y.\
  Berlatzky, J.\ Avron, D.\ Gershoni, B.\ D.\ Gerardot, P.\ M.\ Petroff
  }}(2006)}]{Akopian06}
\bibinfo{author}{\bibnamefont{\rm{N.\ Akopian, N.\ H.\ Lindner, E.\ Poem, Y.\
  Berlatzky, J.\ Avron, D.\ Gershoni, B.\ D.\ Gerardot, P.\ M.\ Petroff }}},
  \bibinfo{journal}{Phys.\ Rev.\ Lett.} \textbf{\bibinfo{volume}{96}},
  \bibinfo{pages}{130501} (\bibinfo{year}{2006}).

\bibitem[{\citenamefont{{F.\ De Martini, G.\ Di Giuseppe, M.\
  Marrocco}}(1996)}]{Martini96}
\bibinfo{author}{\bibnamefont{{F.\ De Martini, G.\ Di Giuseppe, M.\
  Marrocco}}}, \bibinfo{journal}{Phys. Rev. Lett.}
  \textbf{\bibinfo{volume}{76}}, \bibinfo{pages}{900} (\bibinfo{year}{1996}).

\bibitem[{\citenamefont{{P.\ Michler, A.\ Kiraz, C.\ Becher, W.\ V.\
  Schoenfeld, P.\ M.\ Petroff, L.\ Zhang, E.\ Hu, A
  Imamoglu}}(2000)}]{Michler00}
\bibinfo{author}{\bibnamefont{{P.\ Michler, A.\ Kiraz, C.\ Becher, W.\ V.\
  Schoenfeld, P.\ M.\ Petroff, L.\ Zhang, E.\ Hu, A Imamoglu}}},
  \bibinfo{journal}{Science} \textbf{\bibinfo{volume}{290}},
  \bibinfo{pages}{2282} (\bibinfo{year}{2000}).

\bibitem[{\citenamefont{{O.\ Benson, C.\ Santori, M.\ Pelton, Y.\
  Yamamoto}}(2000)}]{Benson00}
\bibinfo{author}{\bibnamefont{{O.\ Benson, C.\ Santori, M.\ Pelton, Y.\
  Yamamoto}}}, \bibinfo{journal}{Phys. Rev. Lett.}
  \textbf{\bibinfo{volume}{84}}, \bibinfo{pages}{2513} (\bibinfo{year}{2000}).

\bibitem[{\citenamefont{{D.\ Fattal, K.\ Inoue, J.\ Vu\v{c}kovi\'{c}, C.\
  Santori, G.\ S.\ Solomon, Y.\ Yamamoto}}(2004)}]{Fattal04}
\bibinfo{author}{\bibnamefont{{D.\ Fattal, K.\ Inoue, J.\ Vu\v{c}kovi\'{c}, C.\
  Santori, G.\ S.\ Solomon, Y.\ Yamamoto}}}, \bibinfo{journal}{Phys. Rev.
  Lett.} \textbf{\bibinfo{volume}{92}}, \bibinfo{pages}{037903}
  (\bibinfo{year}{2004}).

\bibitem[{\citenamefont{{O.\ D.\ D.\ Couto Jr., J.\ Puebla, E.\ A.\ Chekhovich,
  I.\ J.\ Luxmoore, C.\ J.\ Elliott, N.\ Babazadeh, M.\ S.\ Skolnick, A.\ I.\
  Tartakovskii, A.\ B.\ Krysa}}(2011)}]{Couto11}
\bibinfo{author}{\bibnamefont{{O.\ D.\ D.\ Couto Jr., J.\ Puebla, E.\ A.\
  Chekhovich, I.\ J.\ Luxmoore, C.\ J.\ Elliott, N.\ Babazadeh, M.\ S.\
  Skolnick, A.\ I.\ Tartakovskii, A.\ B.\ Krysa}}}, \bibinfo{journal}{Phys.
  Rev. B} \textbf{\bibinfo{volume}{84}}, \bibinfo{pages}{125301}
  (\bibinfo{year}{2011}).

\bibitem[{\citenamefont{{R.\ J.\ Warburton, C.\ Sch\"{a}flein, D.\ Haft, F.\
  Bickel, A.\ Lorke, K.\ Karrai, J.\ M.\ Garcia, W.\ Schoenfeld, P.\ M.\
  Petroff}}(2000)}]{Warburton00}
\bibinfo{author}{\bibnamefont{{R.\ J.\ Warburton, C.\ Sch\"{a}flein, D.\ Haft,
  F.\ Bickel, A.\ Lorke, K.\ Karrai, J.\ M.\ Garcia, W.\ Schoenfeld, P.\ M.\
  Petroff}}}, \bibinfo{journal}{Nature} \textbf{\bibinfo{volume}{405}},
  \bibinfo{pages}{926} (\bibinfo{year}{2000}).

\bibitem[{\citenamefont{{D.\ Heiss, V.\ Jovanov, M.\ Caesar, M.\ Bichler, G.\
  Abstreiter, J.\ J.\ Finley}}(2009)}]{Heiss09}
\bibinfo{author}{\bibnamefont{{D.\ Heiss, V.\ Jovanov, M.\ Caesar, M.\ Bichler,
  G.\ Abstreiter, J.\ J.\ Finley}}}, \bibinfo{journal}{App. Phys. Lett.}
  \textbf{\bibinfo{volume}{94}}, \bibinfo{pages}{072108}
  (\bibinfo{year}{2009}).

\bibitem[{\citenamefont{{A.\ Barenco, M.\ A.\ Dupertuis }}(1995)}]{Beranco95}
\bibinfo{author}{\bibnamefont{{A.\ Barenco, M.\ A.\ Dupertuis }}},
  \bibinfo{journal}{Phys. Rev. B} \textbf{\bibinfo{volume}{52}},
  \bibinfo{pages}{2766} (\bibinfo{year}{1995}).

\bibitem[{\citenamefont{{E.\ Dekel, D.\ Gershoni, E.\ Ehrenfreund, J.\ M.\
  Garcia, P.\ M.\ Petroff}}(2000)}]{Dekel0061}
\bibinfo{author}{\bibnamefont{{E.\ Dekel, D.\ Gershoni, E.\ Ehrenfreund, J.\
  M.\ Garcia, P.\ M.\ Petroff}}}, \bibinfo{journal}{Phys.\ Rev.\ B}
  \textbf{\bibinfo{volume}{61}}, \bibinfo{pages}{11009} (\bibinfo{year}{2000}).

\bibitem[{\citenamefont{{E.\ Poem, Y.\ Kodriano, C.\ Tradonsky, B.\ D.\
  Gerardot, P.\ M.\ Petroff, and D.\ Gershoni}}(2010)}]{Poem10}
\bibinfo{author}{\bibnamefont{{E.\ Poem, Y.\ Kodriano, C.\ Tradonsky, B.\ D.\
  Gerardot, P.\ M.\ Petroff, and D.\ Gershoni}}}, \bibinfo{journal}{Phys.\
  Rev.\ B} \textbf{\bibinfo{volume}{81}}, \bibinfo{pages}{085306}
  (\bibinfo{year}{2010}).

\bibitem[{\citenamefont{{S.\ Alon-Braitbart, E.\ Poem, L.\ Fradkin, N.\
  Akopian, S.\ Vilan, E.\ Lifshitz, E.\ Ehrenfreund, D.\ Gershoni, B.\ D.\
  Gerardot, A.\ Badolato, P.\ M.\ Petroff}}(2006)}]{Braitbart06}
\bibinfo{author}{\bibnamefont{{S.\ Alon-Braitbart, E.\ Poem, L.\ Fradkin, N.\
  Akopian, S.\ Vilan, E.\ Lifshitz, E.\ Ehrenfreund, D.\ Gershoni, B.\ D.\
  Gerardot, A.\ Badolato, P.\ M.\ Petroff}}}, \bibinfo{journal}{Phisica E}
  \textbf{\bibinfo{volume}{32}}, \bibinfo{pages}{127} (\bibinfo{year}{2006}).

\bibitem[{\citenamefont{{E.\ Siebert, T.\ Warming, A.\ Schliwa, E.\ Stock, M.\
  Winkelnkemper, S.\ Rodt, D.\ Bimberg}}(2009)}]{Siebert09}
\bibinfo{author}{\bibnamefont{{E.\ Siebert, T.\ Warming, A.\ Schliwa, E.\
  Stock, M.\ Winkelnkemper, S.\ Rodt, D.\ Bimberg}}}, \bibinfo{journal}{Phys.\
  Rev.\ B} \textbf{\bibinfo{volume}{79}}, \bibinfo{pages}{205321}
  (\bibinfo{year}{2009}).

\bibitem[{\citenamefont{{Y.\ Benny, S.\ Khatsevich, Y.\ Kodriano, E.\ Poem, R.\
  Presman, D.\ Galushko, P.\ M.\ Petroff, D.\ Gershoni}}(2011)}]{Benny10}
\bibinfo{author}{\bibnamefont{{Y.\ Benny, S.\ Khatsevich, Y.\ Kodriano, E.\
  Poem, R.\ Presman, D.\ Galushko, P.\ M.\ Petroff, D.\ Gershoni}}},
  \bibinfo{journal}{Phys. Rev. Lett.} \textbf{\bibinfo{volume}{106}},
  \bibinfo{pages}{040504} (\bibinfo{year}{2011}).

\bibitem[{\citenamefont{{H.\ Benisty, H.\ De Neve, C.\
  Weisbuch}}(1998)}]{Weisbuch98}
\bibinfo{author}{\bibnamefont{{H.\ Benisty, H.\ De Neve, C.\ Weisbuch}}},
  \bibinfo{journal}{IEEE J.\ QE} \textbf{\bibinfo{volume}{34}},
  \bibinfo{pages}{1612} (\bibinfo{year}{1998}).

\bibitem[{\citenamefont{{G.\ Ramon, U.\ Mizrahi, N.\ Akopian, S.\ Braitbart,
  D.\ Gershoni, T.\ L.\ Reinecke, B.\ Gerardot, P.\ M.\
  Petroff}}(2006)}]{Ramon06}
\bibinfo{author}{\bibnamefont{{G.\ Ramon, U.\ Mizrahi, N.\ Akopian, S.\
  Braitbart, D.\ Gershoni, T.\ L.\ Reinecke, B.\ Gerardot, P.\ M.\ Petroff}}},
  \bibinfo{journal}{Phys.\ Rev.\ B} \textbf{\bibinfo{volume}{73}},
  \bibinfo{pages}{205330} (\bibinfo{year}{2006}).

\bibitem[{\citenamefont{\rm{J.\ J.\ Finley, A.\ D.\ Ashmore, A.\ Lema\^{\i}tre,
  D.\ J.\ Mowbray, M.\ S.\ Skolnick, I.\ E.\ Itskevich, P.\ A.\ Maksym, M.\
  Hopkinson, T. F. Krauss}}(2001)}]{Finley01}
\bibinfo{author}{\bibnamefont{\rm{J.\ J.\ Finley, A.\ D.\ Ashmore, A.\
  Lema\^{\i}tre, D.\ J.\ Mowbray, M.\ S.\ Skolnick, I.\ E.\ Itskevich, P.\ A.\
  Maksym, M.\ Hopkinson, T. F. Krauss}}}, \bibinfo{journal}{Phys.\ Rev.\ B}
  \textbf{\bibinfo{volume}{63}}, \bibinfo{pages}{073307}
  (\bibinfo{year}{2001}).

\bibitem[{\citenamefont{{M.\ E.\ Ware, E.\ A.\ Stinaff, D.\ Gammon, M.\ F.\
  Doty, A.\ S.\ Bracker, D.\ Gershoni, V.\ L.\ Korenev, S.\ C.\ Badescu, Y.\
  Lyanda-Geller, T.\ L.\ Reinecke}}(2005)}]{Ware05}
\bibinfo{author}{\bibnamefont{{M.\ E.\ Ware, E.\ A.\ Stinaff, D.\ Gammon, M.\
  F.\ Doty, A.\ S.\ Bracker, D.\ Gershoni, V.\ L.\ Korenev, S.\ C.\ Badescu,
  Y.\ Lyanda-Geller, T.\ L.\ Reinecke}}}, \bibinfo{journal}{Phys.\ Rev.\ Lett.}
  \textbf{\bibinfo{volume}{95}}, \bibinfo{pages}{177403}
  (\bibinfo{year}{2005}).

\bibitem[{\citenamefont{\rm{A.\ Hartmann, Y.\ Ducommun, E.\ Kapon, U.\
  Hohenester, E.\ Molinari}}(2000)}]{Hartmann00}
\bibinfo{author}{\bibnamefont{\rm{A.\ Hartmann, Y.\ Ducommun, E.\ Kapon, U.\
  Hohenester, E.\ Molinari}}}, \bibinfo{journal}{Phys. Rev. Lett.}
  \textbf{\bibinfo{volume}{84}}, \bibinfo{pages}{5648} (\bibinfo{year}{2000}).

\bibitem[{\citenamefont{\rm{D.\ V.\ Regelman, D.\ Gershoni, E.\ Ehrenfreund,
  W.\ V.\ Schoenfeld, P.\ M.\ Petroff}}(2002)}]{Regelman02}
\bibinfo{author}{\bibnamefont{\rm{D.\ V.\ Regelman, D.\ Gershoni, E.\
  Ehrenfreund, W.\ V.\ Schoenfeld, P.\ M.\ Petroff}}}, \bibinfo{journal}{Phys.
  Stat. Sol.} \textbf{\bibinfo{volume}{190}}, \bibinfo{pages}{491}
  (\bibinfo{year}{2002}).

\bibitem[{\citenamefont{{E.\ Poem, Y.\ Kodriano, C.\ Tradonsky, N. \H.\ Lindner,
  B.\ D.\ Gerardot, P.\ M.\ Petroff, D.\ Gershoni}}(2010)}]{Poem102}
\bibinfo{author}{\bibnamefont{{E.\ Poem, Y.\ Kodriano, C.\ Tradonsky, N.\ H.\
  Lindner, B.\ D.\ Gerardot, P.\ M.\ Petroff, D.\ Gershoni}}},
  \bibinfo{journal}{Nat. Phys.} \textbf{\bibinfo{volume}{6}},
  \bibinfo{pages}{993} (\bibinfo{year}{2010}).

\bibitem[{\citenamefont{{V.\ K.\ Kalevich, I.\ A.\ Merkulov, A.\ Yu.\ Shiryaev,
  K.\ V.\ Kavokin, M.\ Ikezawa, T.\ Okuno, P.\ N.\ Brunkov, A.\ E.\ Zhukov, V.\
  M.\ Ustinov, Y.\ Masumoto}}(2005)}]{Kalevich05}
\bibinfo{author}{\bibnamefont{{V.\ K.\ Kalevich, I.\ A.\ Merkulov, A.\ Yu.\
  Shiryaev, K.\ V.\ Kavokin, M.\ Ikezawa, T.\ Okuno, P.\ N.\ Brunkov, A.\ E.\
  Zhukov, V.\ M.\ Ustinov, Y.\ Masumoto}}}, \bibinfo{journal}{Phys.\ Rev.\ B}
  \textbf{\bibinfo{volume}{72}}, \bibinfo{pages}{045325}
  (\bibinfo{year}{2005}).

\bibitem[{\citenamefont{\rm{K.\ V.\ Kavokin}}(2003)}]{Kavokin03}
\bibinfo{author}{\bibnamefont{\rm{K.\ V.\ Kavokin}}}, \bibinfo{journal}{Phys.
  Stat. Sol.} \textbf{\bibinfo{volume}{195}}, \bibinfo{pages}{06157}
  (\bibinfo{year}{2003}).

\bibitem[{\citenamefont{{E.\ L.\ Ivchenko, G.\ E.\ Pikus}}(1997)}]{Ivchenko}
\bibinfo{author}{\bibnamefont{{E.\ L.\ Ivchenko, G.\ E.\ Pikus}}},
  \emph{\bibinfo{title}{Superlattices and other Heterostructures}}
  (\bibinfo{publisher}{Springer-Verlag, Berlin}, \bibinfo{year}{1997}).

\bibitem[{\citenamefont{{D.\ Gammon, E.\ S.\ Snow, B.\ V.\ Shanabrook, D.\ S.\
  Katzer, D.\ Park }}(1996)}]{Gammon96}
\bibinfo{author}{\bibnamefont{{D.\ Gammon, E.\ S.\ Snow, B.\ V.\ Shanabrook,
  D.\ S.\ Katzer, D.\ Park }}}, \bibinfo{journal}{Phys.\ Rev.\ Lett.}
  \textbf{\bibinfo{volume}{76}}, \bibinfo{pages}{3005} (\bibinfo{year}{1996}).

\bibitem[{\citenamefont{{S.\ Hameau, Y.\ Guldner, O.\ Verzelen, R.\ Ferreira,
  G.\ Bastard, J.\ Zeman, A.\ Lema\^{\i}tre, J.M.
  G\'{e}rard}}(1999)}]{Hameau99}
\bibinfo{author}{\bibnamefont{{S.\ Hameau, Y.\ Guldner, O.\ Verzelen, R.\
  Ferreira, G.\ Bastard, J.\ Zeman, A.\ Lema\^{\i}tre, J.M. G\'{e}rard}}},
  \bibinfo{journal}{Phys.\ Rev.\ Lett.} \textbf{\bibinfo{volume}{83}},
  \bibinfo{pages}{4152} (\bibinfo{year}{1999}).

\bibitem[{\citenamefont{{D.\ Sarkar, H.\ P.\ van der Meulen, J.\ M.\ Calleja,
  J.\ M.\ Becker, R.\ J.\ Haug, K.\ Pierz}}(2005)}]{Sarkar05}
\bibinfo{author}{\bibnamefont{{D.\ Sarkar, H.\ P.\ van der Meulen, J.\ M.\
  Calleja, J.\ M.\ Becker, R.\ J.\ Haug, K.\ Pierz}}}, \bibinfo{journal}{Phys.\
  Rev.\ B} \textbf{\bibinfo{volume}{71}}, \bibinfo{pages}{081302}
  (\bibinfo{year}{2005}).

\bibitem[{\citenamefont{{D.\ Sarkar, H.\ P.\ van der Meulen, J.\ M.\ Calleja,
  J.\ M.\ Meyer, R.\ J.\ Haug, K.\ Pierz}}(2008)}]{Sarkar08}
\bibinfo{author}{\bibnamefont{{D.\ Sarkar, H.\ P.\ van der Meulen, J.\ M.\
  Calleja, J.\ M.\ Meyer, R.\ J.\ Haug, K.\ Pierz}}}, \bibinfo{journal}{Appl.\
  Phys.\ Lett.} \textbf{\bibinfo{volume}{92}}, \bibinfo{pages}{181909}
  (\bibinfo{year}{2008}).

\bibitem[{\citenamefont{{A.\ Lema\^{\i}tre, A.\ D.\ Ashmore, J.\ J.\ Finley,
  D.\ J.\ Mowbray, M.\ S.\ Skolnick, M.\ Hopkinson, T.\ F.\
  Krauss}}(2001)}]{Lemaitre01}
\bibinfo{author}{\bibnamefont{{A.\ Lema\^{\i}tre, A.\ D.\ Ashmore, J.\ J.\
  Finley, D.\ J.\ Mowbray, M.\ S.\ Skolnick, M.\ Hopkinson, T.\ F.\ Krauss}}},
  \bibinfo{journal}{Phys.\ Rev.\ B} \textbf{\bibinfo{volume}{63}},
  \bibinfo{pages}{161309} (\bibinfo{year}{2001}).

\bibitem[{\citenamefont{{F.\ Findeis, A.\ Zrenner, G.\ B\"{o}hm, G.\
  Abstreiter}}(2000)}]{Findeis00}
\bibinfo{author}{\bibnamefont{{F.\ Findeis, A.\ Zrenner, G.\ B\"{o}hm, G.\
  Abstreiter}}}, \bibinfo{journal}{Phys.\ Rev.\ B}
  \textbf{\bibinfo{volume}{61}}, \bibinfo{pages}{R10579}
  (\bibinfo{year}{2000}).

\end{thebibliography}
\end{document}